%% file: dNdeta502-preprint.tex
\documentclass[ALICE,manyauthors]{cernphprep} 
\input{commands.tex}  

\usepackage[usenames]{color}
\usepackage{multirow}
\usepackage[normalem]{ulem} 

\RequirePackage{color}\definecolor{RED}{rgb}{1,0,0}\definecolor{BLUE}{rgb}{0,0,1}

\usepackage{fancyhdr}
\usepackage[comma,square,numbers,sort&compress]{natbib}
\usepackage{hyperref}
\usepackage{lineno}

\begin{document}%
\newlength{\figlen}
\setlength{\figlen}{0.75\textwidth}

\begin{titlepage}
\PHyear{2015}
\PHnumber{324}      
\PHdate{14 December}  
%

\title{Centrality dependence of the charged-particle multiplicity\\ density at mid-rapidity in Pb--Pb collisions at \snnbf\ = 5.02 TeV}
\ShortTitle{Centrality dependence of \dNdeta\ in \PbPb\  at \snn\ = 5.02 TeV}   

\Collaboration{ALICE Collaboration\thanks{See Appendix~\ref{app:collab} for the list of collaboration members}}
\ShortAuthor{ALICE Collaboration} 

\begin{abstract}
The pseudorapidity density of charged particles, \dndeta, at
mid-rapidity in \PbPb~collisions has been measured at a center-of-mass
energy per nucleon pair of \snn\ = 5.02 TeV.  For the 5\% most central
collisions we measure a value of \valuecent.  The rise in \dndeta\ as
a function of \snn\ is steeper than that observed in proton--proton
collisions and follows the trend established by measurements at lower
energy.  The increase of \dndeta\ as a function of the average number
of participant nucleons, \avNpart, calculated in a Glauber model, is
compared with the previous measurement at \snn\ = 2.76 TeV.  A constant
factor of about \factorrunIrunII\ describes the increase in \dndeta\
from \snn\ = 2.76  to 5.02 TeV for all centrality
classes, within the measured range of 0--80\% centrality. The
results are also compared to models based on different mechanisms for
particle production in nuclear collisions.
\end{abstract}
\end{titlepage}
\setcounter{page}{2}

\input{dNdeta502-main.tex}               
%
%

\newenvironment{acknowledgement}{\relax}{\relax}
\begin{acknowledgement}
\section*{Acknowledgements}
\input{acknowledgements.tex}    
\end{acknowledgement}

\bibliographystyle{utphys}   
\bibliography{dNdeta502}

\newpage
\appendix
\section{The ALICE Collaboration}
\label{app:collab}
\input{Alice_Authorlist_2015-Dec-13.tex}  
\end{document}

%% file: commands.tex
\newcommand{\MeVc}{MeV/$c$}

\newcommand{\pp}           {pp}
\newcommand{\ppbar}        {\mbox{$\mathrm {p\overline{p}}$}}
\newcommand{\PbPb}         {\mbox{Pb--Pb}}
\newcommand{\pPb}          {\mbox{p--Pb}}
\newcommand{\AuAu}         {\mbox{Au--Au}}
\newcommand{\dAu}          {\mbox{d--Au}}

\newcommand{\dndeta}       {\ensuremath{\mathrm{d}N_\mathrm{ch}/\mathrm{d}\eta}}
\newcommand{\dNdeta}       {\ensuremath{\langle\dndeta\rangle}}

\newcommand{\s}            {\ensuremath{\sqrt{s}}}
\newcommand{\pt}           {\ensuremath{p_{\rm T}}}

\newcommand{\snn}          {\ensuremath{\sqrt{s_{\rm NN}}}}
\newcommand{\snnbf}        {\ensuremath{\mathbf{{\sqrt{s_{\mathbf NN}}}}}}

\newcommand{\Npart}        {\ensuremath{N_\mathrm{part}}}

\newcommand{\avNpart}      {\ensuremath{\langle N_\mathrm{part} \rangle}}

\newcommand{\Ncoll}        {\ensuremath{N_\mathrm{coll}}}

\newcommand{\dNdetape}     {\ensuremath{\frac{2}{\avNpart}\dNdeta}} 

\newcommand{\abs}[1]       {\ensuremath{\left|#1\right|}}

\newcommand{\valuecent} {\ensuremath{1943 \pm 54}}
\newcommand{\factorrunIrunII} {1.2}



%% file: dNdeta502-main.tex
The theory describing the strong interaction, quantum chromodynamics
(QCD), predicts the existence of a deconfined phase of matter, the
quark-gluon plasma, at high temperature and energy density.
Ultra-relativistic collisions of nuclei achieve the conditions
necessary for the formation of this strongly interacting
matter \cite{Karsch:2001cy,Muller:2012zq}.

The multiplicity of produced particles is an important property of the
collisions related to the collision
geometry,  the initial parton densities and the energy density produced. Its dependence on the
impact parameter is sensitive to the interplay between particle
production from hard and soft processes and coherence effects between
individual nucleon--nucleon scatterings. With an increase in the
collision energy, the role of hard processes i.e. parton scatterings
with large momentum transfer, increases. After a two-year long
shutdown, the Large Hadron Collider (LHC) restarted operation in June
2015 and produced \PbPb\ collisions at a per nucleon center-of-mass
energy of \snn\ = 5.02 TeV in November 2015. This is the highest
energy achieved in the laboratory to date and offers the possibility
to further constrain particle production models by studying
their \snn\ dependence.

Collisions of extended objects such as nuclei can be classified
according to their centrality, which is related to the overlap area of
the nuclei. This results in different numbers of nucleons
participating in the collision. The number of these
participants, \Npart, can be calculated by a Monte Carlo (MC) sampling
technique in the Glauber model~\cite{Glaubder.Annurev.Nucl}.

Previous measurements of \dndeta\ for nucleus--nucleus (AA)
collisions were performed at the LHC by
ALICE \cite{PhysRevLett.106.032301}, ATLAS \cite{Aad2012363} and
CMS \cite{Chatrchyan:2011aa} at \snn~= 2.76 TeV and at lower energies,
in the range \snn~= 9 to 200 GeV, with experiments at
the Super Proton Synchrotron (SPS) and Relativistic Heavy Ion Collider
(RHIC) \cite{Abreu:2002fw,Bearden:2001xw,Bearden:2001qq,Adcox:2000sp,Alver:2010ck,Abelev:2008ab}.
They show that the increase of \dndeta\ with energy is steeper in nucleus-nucleus compared to proton--proton collisions.
The centrality dependence of \dNdetape\ in \PbPb\ at \snn~{= 2.76} TeV is very similar to that measured in \snn\ = 200 GeV collisions at RHIC, pointing to a similar mechanism of particle production at the two energies.

In this Letter we present the measurement of the charged-particle
pseudorapidity density averaged in the interval
$\abs{\eta}<0.5$, \dNdeta, and its centrality
dependence.  The pseudorapidity is defined by $\eta \equiv~{
- \ln{\tan{(\theta/2)}}}$, with $\theta$ the emission angle of the
particle relative to the beam axis. The primary charged particles are
defined as prompt particles produced in the collision including all
decay products, except products from weak decays of light flavor
hadrons and of muons.

The data were recorded with the ALICE detector in November 2015
at \snn\ = 5.02 TeV. Full details on the ALICE
apparatus \cite{Aamodt:2008zz} and its operational
performance \cite{Abelev:2014ffa} are given elsewhere. A brief
description of the most relevant elements, along with the experimental
conditions, follows. The observed interaction rate was around 300 Hz,
of which about 25 Hz were from hadronic interactions, the remainder
being a background from electromagnetically induced processes. A total
of about $10^5$ hadronic events are used. The interaction probability
per bunch-crossing (during which bunches of ions from each beam are
arranged to be co-incident at the ALICE interaction point) was
sufficiently small that the chance of two hadronic interactions
occurring together, so-called pileup events, was negligible.

The measurement relies on the ALICE Inner Tracking System, the
innermost two layers of which form the Silicon Pixel Detector (SPD). 
It consists of arrays of pixels arranged with an approximate
cylindrical geometry at radii of 3.9 and 7.6 cm covering intervals of
$\left|\eta\right| < 2.0$ and $\left|\eta\right| < 1.4$ for the inner
and outer layers, respectively.  The SPD is situated in a solenoidal
magnet, with its principal axis along the beam line, providing a 0.5 T
magnetic field.  The interaction trigger is provided by two detectors,
V0A and V0C, which consist of arrays of scintillators, covering the
full azimuth and more than 4 units of pseudorapidity, in the ranges
$2.8 < \eta < 5.1$ and $-3.7 < \eta < -1.7$, respectively. In all
cases the $\eta$-coverage refers to collisions at the nominal
interaction point. A signal must be present in both V0 detectors to
trigger the recording of the interaction. The V0 detectors also
provide a signal proportional to the number of charged particles
striking them which is used to classify the events into centrality
classes, defined in terms of percentiles of the hadronic
cross-section. In addition, an offline event selection employs the
information from two Zero Degree Calorimeters (ZDC) positioned 112.5 m
from the interaction point on either side. Beam background events are
removed using the V0 timing information and the correlation between
the sum and the difference of times measured in each of the ZDCs \cite{Abelev:2014ffa}.

The analysis is restricted to the 80\% most central events.
The classification of events into centrality classes is done by
using the summed amplitudes of the signals in the V0A and V0C
detectors, following the method developed previously
\cite{ALICE-PUBLIC-2015-008,PhysRevC.88.044909}. The V0 amplitude
is fitted with an MC implementation of the Glauber model coupled with a two-component model
assuming that the effective number of particle-producing sources is
given by ${f \times \Npart + (1-f) \times \Ncoll}$, where \Npart\ is
the number of participating nucleons, \Ncoll\ is the number of binary
nucleon--nucleon collisions and $f \sim 0.8$ quantifies their relative
contributions. The number of particles produced by each source is
distributed according to a Negative Binomial Distribution (NBD),
parametrised with $\mu$ and $k$, where $\mu$ is the mean multiplicity
per source and $k$ controls the contribution at high multiplicity.  In
the Monte Carlo Glauber calculation, the nuclear density for
$^{208}$Pb is modeled by a Woods--Saxon distribution for a spherical
nucleus with a radius of $6.62 \pm 0.06$ fm and a skin thickness of
$0.546 \pm 0.010$ fm, based on data from low energy electron--nucleus
scattering experiments \cite{DEVRIES1987495}, and a hard-sphere
exclusion distance between nucleons of $0.4 \pm 0.4$ fm. For \snn\ =
5.02 TeV collisions, an inelastic nucleon--nucleon cross-section of
$70 \pm 5$ mb, obtained by interpolation \cite{ALICE:2012xs}, is used.
The fit was restricted to a region where the effects of trigger inefficiency and
contamination by electromagnetic processes are negligible.
The NBD-Glauber fit provides a good description of the
observed V0 amplitude in this region, which corresponds to the most
central 90\% of the cross-section. All events in the sample corresponding to 0--80\%
of the hadronic cross section are found to have a well-defined primary
vertex, extracted by correlating hits in the two SPD layers.

The \dndeta\ measurement is performed using short track segments,
termed tracklets \cite{Aamodt:2010pb}.  Tracklet candidates are formed
using the position of the primary vertex and a pair of hits, one in
each SPD layer.  
For each of the hits in the pair two angles are determined with
respect to the reconstructed interaction vertex and the angular
differences, $\Delta{\varphi}$ in the bending plane and
$\Delta{\theta}$ in the polar direction, are calculated for each pair
of hits. In order to reject candidates produced by the random
combination of two hits, tracklets are selected by a cut on the sum of
the squares, $\delta^{2} = (\Delta{\varphi}/\sigma_{\varphi})^2 +
(\Delta{\theta}/\sigma_{\theta})^2 < 1.5$, where $\sigma_{\varphi} =
60$ mrad and $\sigma_{\theta} = 25\sin^2\theta$ mrad.
This selection effectively allows the reconstruction of
charged particles with transverse momentum (\pt) above the 50 MeV/c
cut-off determined by particle absorption in the material.

The acceptance region in $\eta$ depends on the position of the
interaction vertex along the beamline, $z$. Events with $\abs{z}<7$ cm are
used, corresponding to a coverage of $\abs{\eta}<0.5$ with an
approximately constant acceptance.

A correction is needed to account for
the acceptance and efficiency of a primary track
to generate a tracklet, including the extrapolation to zero \pt, and
for the removal of combinatorial background tracklets. This is
computed using simulated data from the \textsc{HIJING} event
generator \cite{Wang:1991hta} transported through
a \textsc{GEANT3} \cite{GEANT3} simulation of ALICE, where the centrality definition is adjusted so that the particle density is similar to that in real data for the same centrality classes. A re-weighting of
the generator output is performed to reproduce the \pt\ distributions
of inclusive charged hadrons and the relative abundances of pions,
protons, kaons and other strange particles as measured in \PbPb\
collisions
at \snn~=~2.76~TeV \cite{Abelev:2012eq,PhysRevC.88.044910,Abelev:2013xaa,ABELEV:2013zaa}. Using
results from \snn\ = 2.76 TeV is justified because the relative
abundances at ${\snn\ = 2.76}$ TeV change very little from those
at \snn\ = 200 GeV. Any variation with the increase in \snn\ to 5.02
TeV will be much smaller than the differences between the default and
re-weighted \textsc{HIJING} simulations, which lead to differences in
the results within the systematic uncertainties estimated below.

The correction takes into account any inactive channels present at the
time of data taking as well as losses due to physical processes like
absorption and scattering, which may result in a charged particle not
creating a tracklet. The fractions of active pixels in the inner
and outer SPD layers were about 85\% and 97.5\%, respectively. The
estimated combinatorial background amounts to about 18\% in the most
central (0--2.5\%) and 1\% in the most peripheral (70--80\%)
centrality classes. A correction of about 2\% for contamination by
secondaries from weak decays is applied based on the same simulation.

\begin{table}[t]
\centering
\begin{tabular}{@{} c|c|c|c @{}} 
Centrality & \dNdeta  & \avNpart  & \dNdetape \\
\hline
0--2.5\%   & 2035 $\pm$ 52  & 398  $\pm$ 2    & 10.2 $\pm$ 0.3 \\
2.5--5.0\% & 1850 $\pm$ 55  & 372  $\pm$ 3    & 9.9  $\pm$ 0.3 \\
5.0--7.5\% & 1666 $\pm$ 48  & 346  $\pm$ 4    & 9.6  $\pm$ 0.3 \\
7.5--10\%  & 1505 $\pm$ 44  & 320  $\pm$ 4    & 9.4  $\pm$ 0.3 \\
10--20\%   & 1180 $\pm$ 31  & 263  $\pm$ 4    & 9.0  $\pm$ 0.3 \\
20--30\%   & 786  $\pm$ 20  & 188  $\pm$ 3    & 8.4  $\pm$ 0.3 \\
30--40\%   & 512  $\pm$ 15  & 131  $\pm$ 2    & 7.8  $\pm$ 0.3 \\
40--50\%   & 318  $\pm$ 12  & 86.3 $\pm$ 1.7  & 7.4  $\pm$ 0.3 \\
50--60\%   & 183  $\pm$ 8   & 53.6 $\pm$ 1.2  & 6.8  $\pm$ 0.3 \\
60--70\%   & 96.3 $\pm$ 5.8 & 30.4 $\pm$ 0.8  & 6.3  $\pm$ 0.4 \\
70--80\%   & 44.9 $\pm$ 3.4 & 15.6 $\pm$ 0.5  & 5.8  $\pm$ 0.5 \\
\end{tabular}
\caption{\label{tab:data}
The $\dNdeta$ and $\dNdetape$ values measured in $\abs{\eta}<0.5$ for eleven centrality classes.
The values of $\avNpart$ obtained with the Glauber model are also given. The errors are total uncertainties, the statistical contribution being negligible.} 
\end{table}

\begin{figure}[t]
\begin{center}
\includegraphics[width=\figlen]{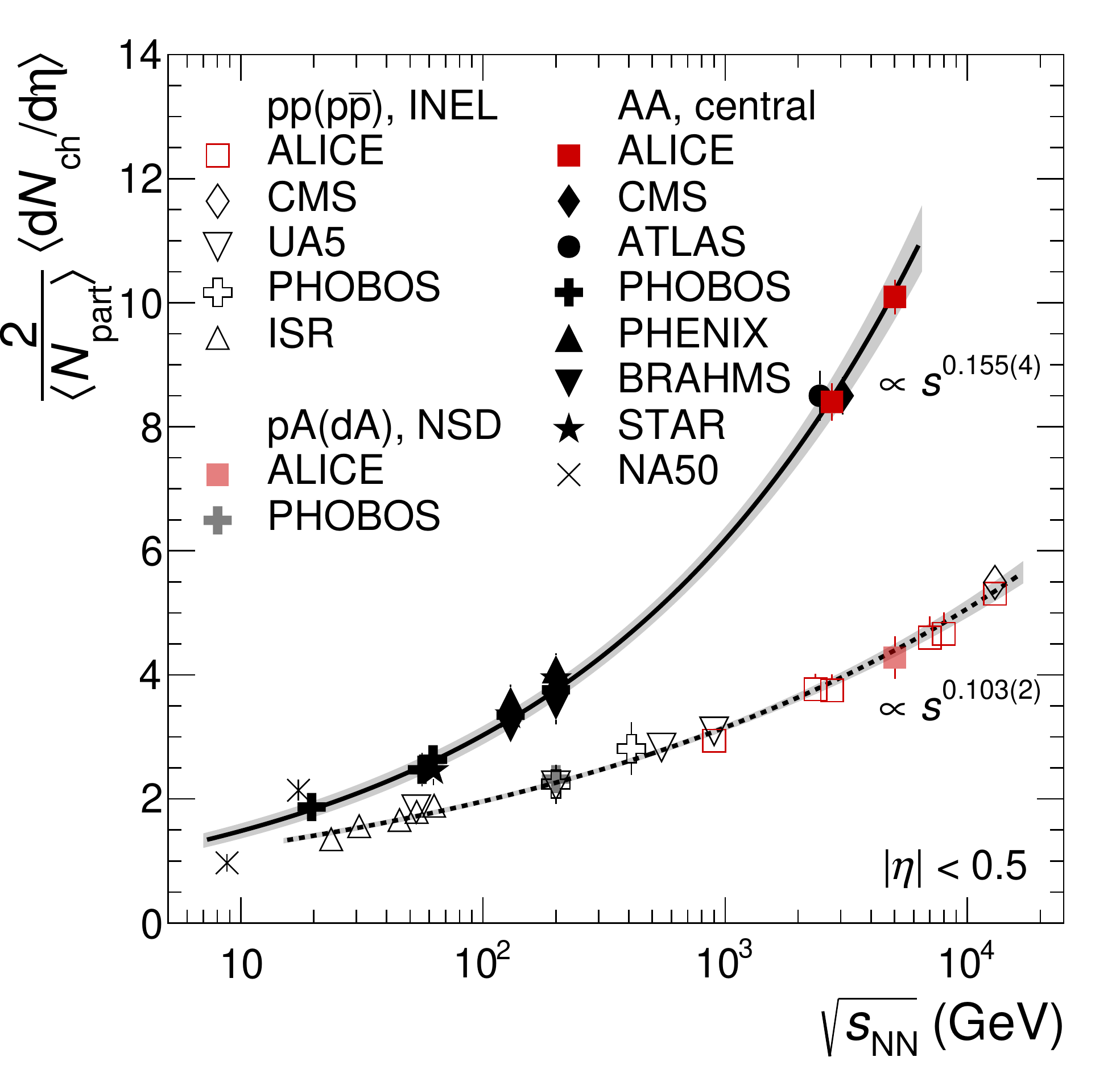}
\caption{Values of $\dNdetape$ for central \PbPb\ \cite{PhysRevLett.106.032301,Aad2012363,Chatrchyan:2011aa,Abreu:2002fw} and \AuAu\ \cite{Bearden:2001xw,Bearden:2001qq,Adcox:2000sp,Alver:2010ck,Abelev:2008ab} collisions (see text) as a function of \snn. Measurements for inelastic \pp\ and \ppbar\ collisions as a function of \s\ are also shown \cite{LongMultiPaper,Khachatryan2015143,Adam2016319} along with those from non-single diffractive p--A and d--A collisions \cite{PhysRevLett.110.032301,Back:2003hx}. 
The $s$-dependencies of the AA and \pp\ (\ppbar) collision data are well described by the functions $s_{\rm NN}^{0.155}$ (solid line) and $s_{\rm NN}^{0.103}$ (dashed line), respectively.
The shaded bands show the uncertainties on the extracted power-law
dependencies. The central \PbPb\ measurements from CMS and ATLAS at
2.76 TeV have been shifted horizontally for clarity.}
\label{fig:roots}
\end{center}
\end{figure}

\begin{figure}[t]
\begin{center}
\includegraphics[width=\figlen]{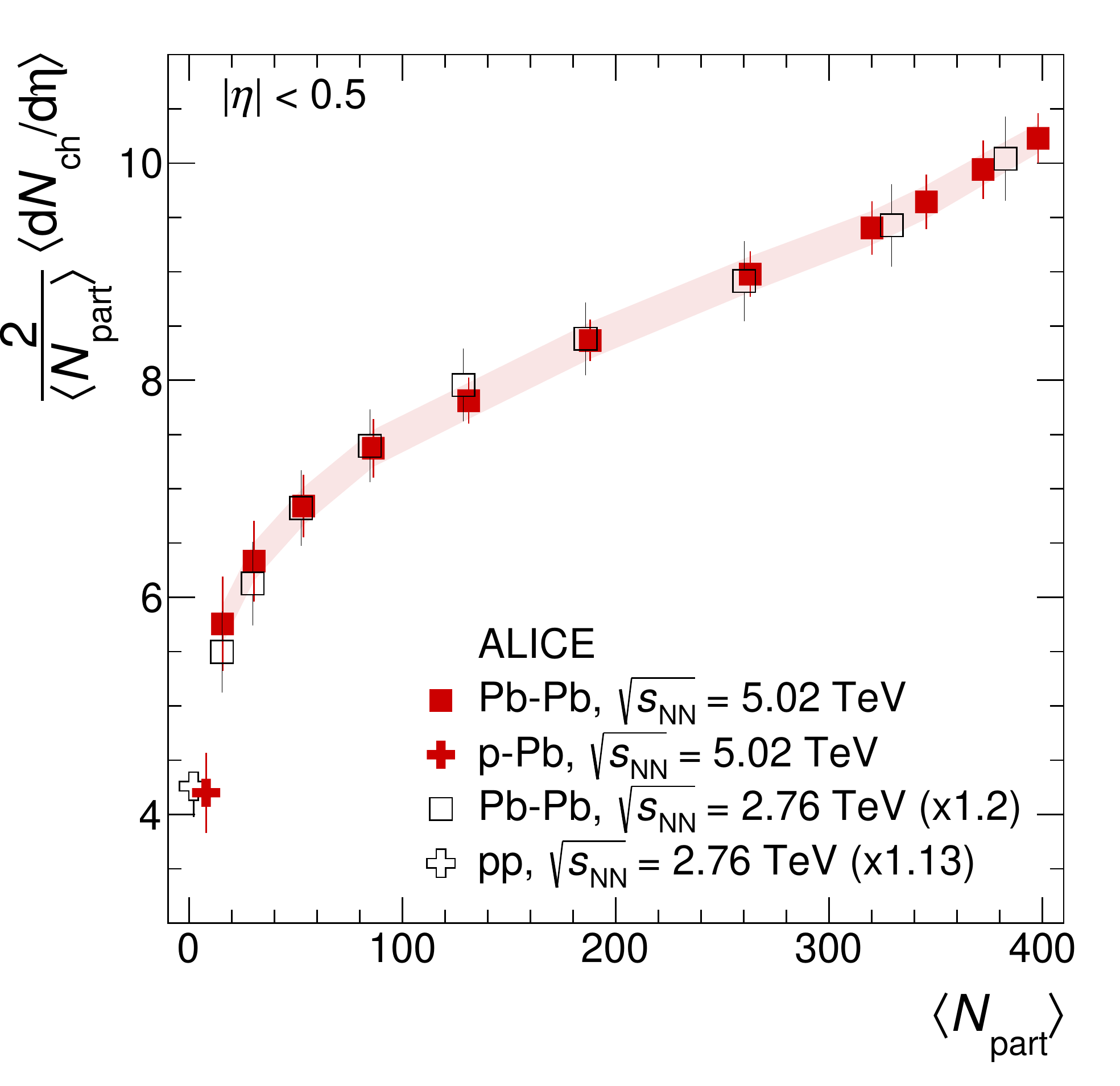}
\caption{The $\dNdetape$ for \PbPb\ collisions at \snn\ =~5.02 TeV in the centrality range 0--80\%, as a function of $\langle N_{\mathrm{part}} \rangle$ in each centrality class. The error bars indicate the point-to-point centrality-dependent uncertainties whereas the shaded band shows the correlated contributions. Also shown is the result from non-single diffractive \pPb\ collisions at the same \snn\ \cite{ALICE:2012xs}. Data from lower energy (2.76 TeV) \PbPb\  and pp collisions \cite{PhysRevLett.106.032301,LongMultiPaper}, scaled by a factor of 1.2 and 1.13 respectively, are shown for comparison. The error bars for \pPb\ at \snn\ = 5.02 TeV and lower energy \PbPb\ and pp collisions indicate the total uncertainty.}

\label{fig:npart}
\end{center}
\end{figure}

\begin{figure}[t]
\begin{center}
\includegraphics[width=\figlen]{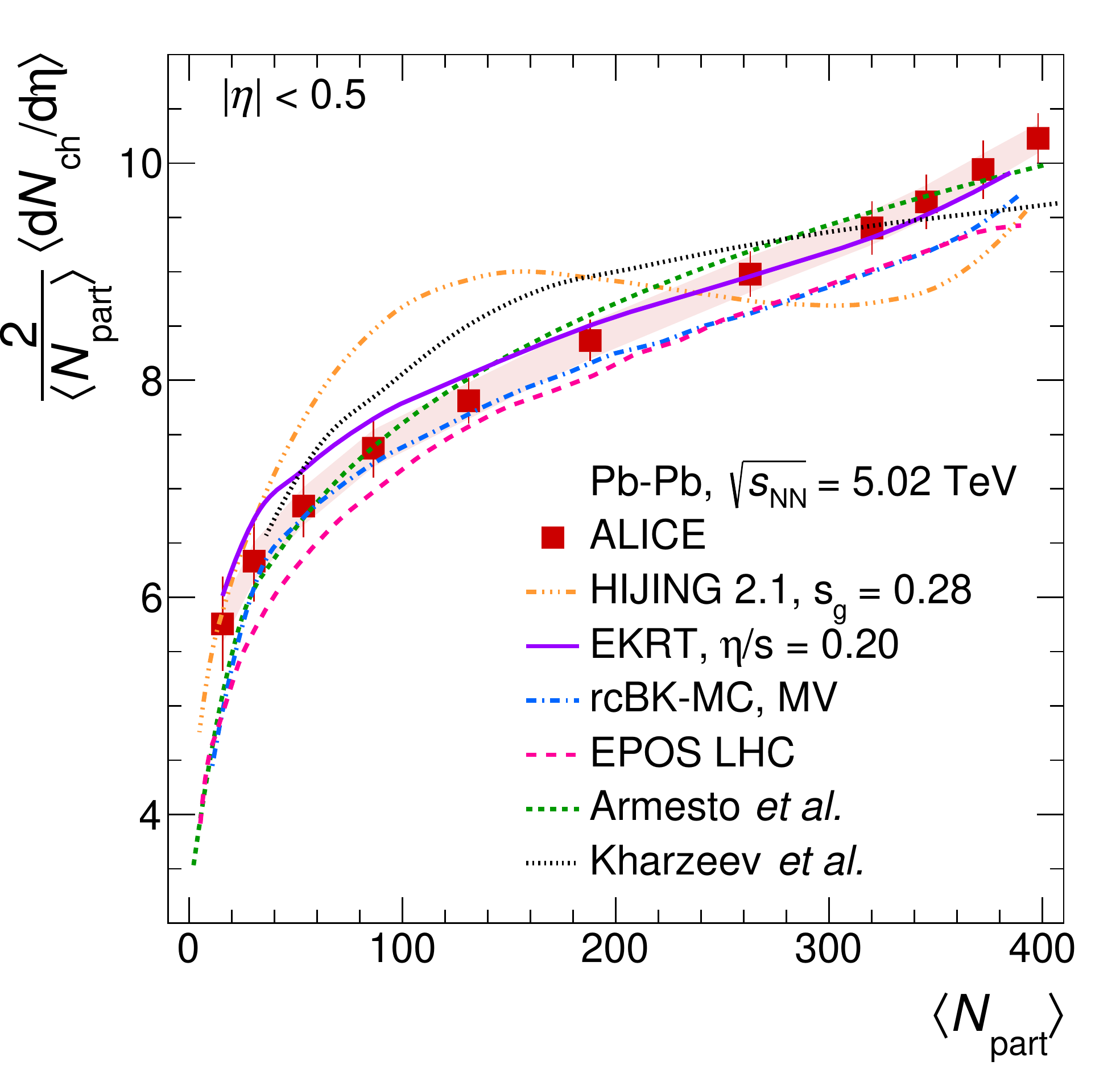}
\caption{The $\dNdetape$ for \PbPb\ collisions at \snn\ = 5.02 TeV in the centrality range 0-80\%, as function a of $\langle N_{\mathrm{part}} \rangle$ in each centrality class, compared to model predictions \cite{Niemi:2015qia,Niemi:2015voa,PhysRevC.83.014915,PhysRevC.85.044920,Albacete:2011fw,Albacete:2010ad,Armesto:2004ud,Kharzeev:2004if,PhysRevC.92.034906}.}
\label{fig:models}
\end{center}
\end{figure}

Several sources of systematic uncertainty were investigated. The
centrality determination introduces an uncertainty via the fitting of
the V0 amplitude distribution to the hadronic cross-section, due to
the contamination from electromagnetically induced reactions at small
multiplicity.  The fraction of the hadronic cross-section (10\%)
at the lowest multiplicity, where the trigger and event
selection are not fully efficient and the contamination is
non-negligible, was varied by an uncertainty of $\pm 0.5\%$. This uncertainty was
estimated by varying NBD-Glauber fitting conditions and by
fitting a different centrality estimator, based on the hits in the
SPD. The uncertainty from the centrality estimation results in an
uncertainty of 0.5\% for central 0--2.5\% collisions,
increasing in the more peripheral collision classes,
reaching 7.5\% for the 70--80\% sample,
where it is the largest contribution. Conversely, the uncertainty due
to the subtraction of the background is largest for the central event
sample, where it is about 2\%, and becomes smaller as the collisions
become more peripheral, amounting to only 0.2\% for the 70--80\% event
class. This uncertainty is estimated by using an alternative method where fake hits
are injected into real events.

All other sources of systematic uncertainty are independent of
centrality. The uncertainty resulting from the subtraction of the
contamination from weak decays of strange hadrons is estimated, from
the tuned MC simulations, to amount to about 0.5\% by varying the
strangeness content by $\pm$30\%.  The uncertainty due to the
extrapolation down to zero \pt\ is estimated to be about 0.5\% by
varying the number of particles below the 50 \MeVc\ low-\pt\ cut-off
by $\pm$30\%.  An uncertainty of 1\% for variations in detector
acceptance and efficiency was evaluated by carrying out the analysis
for different slices of the $z$-position of the interaction vertex
distribution and with subsamples in azimuth.

Other effects due to particle composition, background events, pileup,
material budget and tracklet selection criteria were found to be
negligible. The final systematic uncertainties assigned to the
measurements are the quadratic sums of the individual contributions,
and range from 2.6\% in central 0--2.5\% collisions to 7.6\% in
70--80\% peripheral collisions, of which 2.3\% and 7.5\%,
respectively, are centrality dependent and 1.2\% are centrality
independent.
 
The results for \dNdeta\ are shown in Table \ref{tab:data}.
In order to compare bulk particle production at different energies and in different collision systems, specifically for a direct comparison to \pp\ and \ppbar\ collisions,
the charged-particle density is divided by the average number of
participating nucleon pairs, \avNpart/2.
The \avNpart\ values are calculated with an MC-Glauber  for centrality classes defined by classifying the events according to their impact parameter and are also listed in Table \ref{tab:data}. The systematic uncertainty on \avNpart\ is obtained by independently varying the parameters of the Glauber model within their estimated uncertainties.
For the most central 0--5\% collisions, a density of primary charged particles at mid-rapidity \dNdeta\ = \valuecent\ was measured and, normalized per participant pair corresponds to \dNdetape\ = 10.1 $\pm$ 0.3.
In Figure 1 this value is compared to the existing data for
central \PbPb\ and \AuAu\ collisions from experiments at
LHC \cite{PhysRevLett.106.032301,Aad2012363,Chatrchyan:2011aa},
RHIC \cite{Bearden:2001xw,Bearden:2001qq,Adcox:2000sp,Alver:2010ck,Abelev:2008ab}
and SPS \cite{Abreu:2002fw}. The data shown are for 0--5\% except for
the results from PHOBOS \cite{Alver:2010ck} and
ATLAS \cite{Aad2012363} which are for 0--6\%. The dependence
of \dNdetape\ on the center-of-mass energy can be fitted with a power
law of the form $a\cdot s^{b}$. This gives an
exponent, under the assumption of uncorrelated uncertainties, of $b =
0.155 \pm 0.004$. It is a much stronger $s$-dependence than for
proton--proton collisions, where a value of $b = 0.103 \pm 0.002$ is
obtained from a fit to the same function \cite{Adam2016319}. The fit
results are plotted with their uncertainties shown as shaded
bands. The result at \snn~=~5.02~TeV confirms the trend established by
lower energy data since $b$ is not significantly different when the
new point is excluded from the fit. It can also be seen in the figure
that the values of \dNdetape\ measured by ALICE
for \pPb\ \cite{ALICE:2012xs} and PHOBOS for \dAu\
\cite{Alver:2010ck} collisions fall on the curve for proton--proton collisions, indicating that the strong rise in AA is not solely related to the multiple collisions undergone by the participants since the proton in p--A collisions also encounters multiple nucleons.

The centrality dependence of \dNdetape\ is shown in
Figure \ref{fig:npart}.  The point-to-point centrality-dependent
uncertaintes are indicated by error bars whereas the shaded bands show
the correlated contributions. The statistical uncertainties are
negligible. The data are plotted as a function of \avNpart\ and a
strong dependence is observed, with \dNdetape\ decreasing by a factor
1.8 from the most central collisions, large \avNpart, to the most
peripheral, small \avNpart. There appears to be a smooth trend towards
the value measured in minimum bias \pPb\
collisions \cite{ALICE:2012xs}. The \PbPb\ data measured at \snn\ = 2.76
TeV \cite{PhysRevLett.106.032301} are also shown,
scaled by a factor 1.2, which is calculated from the observed
$s^{0.155}$ dependence of the results in the most central collisions,
and which describes well the increase for all centralities. The proton--proton result at the same energy \cite{LongMultiPaper} is scaled by a factor 1.13 from the $s^{0.103}$ dependence. The ratio
between the data measured at the two collision energies is consistent
with being independent of \Npart, within the uncertainties, which are
largely uncorrelated. While in general the uncertainties related to
the tracklet measurement are correlated between the two analyses, the
subtraction of the background and the centrality classification are, instead, uncorrelated, depending on the determination of the usable fraction of the hadronic
cross-section and therefore on the run and detector conditions \cite{ALICE-PUBLIC-2015-008}.

Figure \ref{fig:models} shows a comparison of the data to some of the
models which were compared to the measurements at lower energy. The
curves shown are predictions of the models, without any retuning of
the parameters based on the new data presented here.

Predictions from commonly used Monte Carlo
generators, \textsc{HIJING} \cite{PhysRevC.83.014915} and \textsc{EPOS
LHC} \cite{PhysRevC.92.034906}, are also shown.  HIJING combines
perturbative-QCD (pQCD) processes with soft interactions, and includes
a strong impact parameter dependence of parton shadowing. The data
at \snn\ = 2.76 TeV were previously compared to \textsc{HIJING} using
gluon shadowing parameter, $s_{g}$, values of 0.20 and
0.23 \cite{PhysRevLett.106.032301}. The higher value gave a better
estimate of the overall normalization, the lower one a better
agreement with the shape. At \snn\ = 5.02 TeV a larger $s_{g}$ value
of 0.28 is required to limit the multiplicity per participant, leading
to a centrality dependence which does not reproduce the data. EPOS is
a model based on the Gribov-Regge theory at parton level which
incorporates collective effects treated via a flow parametrisation in
the EPOS LHC version. It provides a good description of the data.

Saturation-inspired models
(rcBK-MC, with the MV initial conditions \cite{Albacete:2011fw,Albacete:2010ad}, Kharzeev \emph{et
al.} \cite{Kharzeev:2004if} and Armesto \emph{et
al.} \cite{Armesto:2004ud}) rely on pQCD and use an initial-state
gluon density to fix an energy-dependent scale at which the quark and
gluon densities saturate thereby limiting the number of produced
partons and, in turn, of particles. This results in a factorization of
the energy and centrality dependences of the multiplicity in the
models, as observed in the experimental data.  The rcBK-MC and
Armesto \emph{et al.} models provide a better description of the data, in particular
of the shape, than the Kharzeev \emph{et al.} model. 

The EKRT model \cite{Niemi:2015qia,Niemi:2015voa} combines collinearly
factorized next-to-leading order pQCD mini-jet cross sections with a
conjecture of gluon saturation to suppress soft parton
production. Impact-parameter dependent EPS09s parton distribution
functions \cite{Helenius:2012wd} are used. The space-time evolution of
the system with the computed initial conditions is described with
relativistic viscous hydrodynamics event-by-event. The normalization
is fixed by exploiting the 0--5\% most central multiplicity
measurement \cite{Aamodt:2010pb}. The EKRT model can broadly describe
both the shape and the overall magnitude of the dependence of
multiplicity on centrality.  In general, theoretical models need some
sort of mechanism to limit the growth of multiplicity in order to
describe the centrality and energy evolution of the multiplicity.

In summary, we have measured the charged-particle pseudorapidity
density \dNdeta\ in \PbPb\ collisions at the highest available
center-of-mass energy and observe a 20\% increase for the most
central collisions with respect to similar measurements at 2.76 TeV,
in agreement with the previously established power-law dependence of
this quantity.  The centrality dependence of \dndeta\ is very similar
to that previously measured in lower energy AA collisions, with a
factor of 1.8 increase from peripheral to central collisions. Most of
the models which were able to reproduce the data at \snn~=~2.76~TeV
are able to describe the data at \snn~=~5.02~TeV.  Our results provide
further constraints for models describing high-energy heavy-ion
collisions.

%% file: acknowledgements.tex

The ALICE Collaboration would like to thank all its engineers and technicians for their invaluable contributions to the construction of the experiment and the CERN accelerator teams for the outstanding performance of the LHC complex.
The ALICE Collaboration gratefully acknowledges the resources and support provided by all Grid centres and the Worldwide LHC Computing Grid (WLCG) collaboration.
The ALICE Collaboration acknowledges the following funding agencies for their support in building and
running the ALICE detector:
State Committee of Science,  World Federation of Scientists (WFS)
and Swiss Fonds Kidagan, Armenia;
Conselho Nacional de Desenvolvimento Cient\'{\i}fico e Tecnol\'{o}gico (CNPq), Financiadora de Estudos e Projetos (FINEP),
Funda\c{c}\~{a}o de Amparo \`{a} Pesquisa do Estado de S\~{a}o Paulo (FAPESP);
National Natural Science Foundation of China (NSFC), the Chinese Ministry of Education (CMOE)
and the Ministry of Science and Technology of China (MSTC);
Ministry of Education and Youth of the Czech Republic;
Danish Natural Science Research Council, the Carlsberg Foundation and the Danish National Research Foundation;
The European Research Council under the European Community's Seventh Framework Programme;
Helsinki Institute of Physics and the Academy of Finland;
French CNRS-IN2P3, the `Region Pays de Loire', `Region Alsace', `Region Auvergne' and CEA, France;
German Bundesministerium fur Bildung, Wissenschaft, Forschung und Technologie (BMBF) and the Helmholtz Association;
General Secretariat for Research and Technology, Ministry of Development, Greece;
National Research, Development and Innovation Office (NKFIH), Hungary;
Department of Atomic Energy and Department of Science and Technology of the Government of India;
Istituto Nazionale di Fisica Nucleare (INFN) and Centro Fermi -
Museo Storico della Fisica e Centro Studi e Ricerche ``Enrico Fermi'', Italy;
Japan Society for the Promotion of Science (JSPS) KAKENHI and MEXT, Japan;
Joint Institute for Nuclear Research, Dubna;
National Research Foundation of Korea (NRF);
Consejo Nacional de Cienca y Tecnologia (CONACYT), Direccion General de Asuntos del Personal Academico(DGAPA), M\'{e}xico, Amerique Latine Formation academique - 
European Commission~(ALFA-EC) and the EPLANET Program~(European Particle Physics Latin American Network);
Stichting voor Fundamenteel Onderzoek der Materie (FOM) and the Nederlandse Organisatie voor Wetenschappelijk Onderzoek (NWO), Netherlands;
Research Council of Norway (NFR);
National Science Centre, Poland;
Ministry of National Education/Institute for Atomic Physics and National Council of Scientific Research in Higher Education~(CNCSI-UEFISCDI), Romania;
Ministry of Education and Science of Russian Federation, Russian
Academy of Sciences, Russian Federal Agency of Atomic Energy,
Russian Federal Agency for Science and Innovations and The Russian
Foundation for Basic Research;
Ministry of Education of Slovakia;
Department of Science and Technology, South Africa;
Centro de Investigaciones Energeticas, Medioambientales y Tecnologicas (CIEMAT), E-Infrastructure shared between Europe and Latin America (EELA), 
Ministerio de Econom\'{i}a y Competitividad (MINECO) of Spain, Xunta de Galicia (Conseller\'{\i}a de Educaci\'{o}n),
Centro de Aplicaciones Tecnológicas y Desarrollo Nuclear (CEA\-DEN), Cubaenerg\'{\i}a, Cuba, and IAEA (International Atomic Energy Agency);
Swedish Research Council (VR) and Knut $\&$ Alice Wallenberg
Foundation (KAW);
Ukraine Ministry of Education and Science;
United Kingdom Science and Technology Facilities Council (STFC);
The United States Department of Energy, the United States National
Science Foundation, the State of Texas, and the State of Ohio;
Ministry of Science, Education and Sports of Croatia and  Unity through Knowledge Fund, Croatia;
Council of Scientific and Industrial Research (CSIR), New Delhi, India;
Pontificia Universidad Cat\'{o}lica del Per\'{u}.

%% file: Alice_Authorlist_2015-Dec-13.tex


\begingroup
\small
\begin{flushleft}
J.~Adam\Irefn{org40}\And
D.~Adamov\'{a}\Irefn{org84}\And
M.M.~Aggarwal\Irefn{org88}\And
G.~Aglieri Rinella\Irefn{org36}\And
M.~Agnello\Irefn{org110}\And
N.~Agrawal\Irefn{org48}\And
Z.~Ahammed\Irefn{org132}\And
S.~Ahmad\Irefn{org19}\And
S.U.~Ahn\Irefn{org68}\And
S.~Aiola\Irefn{org136}
% ...
\renewcommand\labelenumi{\textsuperscript{\theenumi}~}

\section*{Affiliation notes}
\renewcommand\theenumi{\roman{enumi}}
\begin{Authlist}
\item \Adef{0}Deceased
\item \Adef{idp1762240}{Also at: Georgia State University, Atlanta, Georgia, United States}
\item \Adef{idp3123728}{Also at: Also at Department of Applied Physics, Aligarh Muslim University, Aligarh, India}
\item \Adef{idp3830352}{Also at: M.V. Lomonosov Moscow State University, D.V. Skobeltsyn Institute of Nuclear, Physics, Moscow, Russia}
\end{Authlist}

\section*{Collaboration Institutes}
\renewcommand\theenumi{\arabic{enumi}~}
\begin{Authlist}

\item \Idef{org1}A.I. Alikhanyan National Science Laboratory (Yerevan Physics Institute) Foundation, Yerevan, Armenia
\item \Idef{org2}Benem\'{e}rita Universidad Aut\'{o}noma de Puebla, Puebla, Mexico
\item \Idef{org3}Bogolyubov Institute for Theoretical Physics, Kiev, Ukraine
\item \Idef{org4}Bose Institute, Department of Physics and Centre for Astroparticle Physics and Space Science (CAPSS), Kolkata, India
\item \Idef{org5}Budker Institute for Nuclear Physics, Novosibirsk, Russia
\item \Idef{org6}California Polytechnic State University, San Luis Obispo, California, United States
\item \Idef{org7}Central China Normal University, Wuhan, China
\item \Idef{org8}Centre de Calcul de l'IN2P3, Villeurbanne, France
\item \Idef{org9}Centro de Aplicaciones Tecnol\'{o}gicas y Desarrollo Nuclear (CEADEN), Havana, Cuba
\item \Idef{org10}Centro de Investigaciones Energ\'{e}ticas Medioambientales y Tecnol\'{o}gicas (CIEMAT), Madrid, Spain
\item \Idef{org11}Centro de Investigaci\'{o}n y de Estudios Avanzados (CINVESTAV), Mexico City and M\'{e}rida, Mexico
\item \Idef{org12}Centro Fermi - Museo Storico della Fisica e Centro Studi e Ricerche ``Enrico Fermi'', Rome, Italy
\item \Idef{org13}Chicago State University, Chicago, Illinois, USA
\item \Idef{org14}China Institute of Atomic Energy, Beijing, China
\item \Idef{org15}Commissariat \`{a} l'Energie Atomique, IRFU, Saclay, France
\item \Idef{org16}COMSATS Institute of Information Technology (CIIT), Islamabad, Pakistan
\item \Idef{org17}Departamento de F\'{\i}sica de Part\'{\i}culas and IGFAE, Universidad de Santiago de Compostela, Santiago de Compostela, Spain
\item \Idef{org18}Department of Physics and Technology, University of Bergen, Bergen, Norway
\item \Idef{org19}Department of Physics, Aligarh Muslim University, Aligarh, India
\item \Idef{org20}Department of Physics, Ohio State University, Columbus, Ohio, United States
\item \Idef{org21}Department of Physics, Sejong University, Seoul, South Korea
\item \Idef{org22}Department of Physics, University of Oslo, Oslo, Norway
\item \Idef{org23}Dipartimento di Elettrotecnica ed Elettronica del Politecnico, Bari, Italy
\item \Idef{org24}Dipartimento di Fisica dell'Universit\`{a} 'La Sapienza' and Sezione INFN Rome, Italy
\item \Idef{org25}Dipartimento di Fisica dell'Universit\`{a} and Sezione INFN, Cagliari, Italy
\item \Idef{org26}Dipartimento di Fisica dell'Universit\`{a} and Sezione INFN, Trieste, Italy
\item \Idef{org27}Dipartimento di Fisica dell'Universit\`{a} and Sezione INFN, Turin, Italy
\item \Idef{org28}Dipartimento di Fisica e Astronomia dell'Universit\`{a} and Sezione INFN, Bologna, Italy
\item \Idef{org29}Dipartimento di Fisica e Astronomia dell'Universit\`{a} and Sezione INFN, Catania, Italy
\item \Idef{org30}Dipartimento di Fisica e Astronomia dell'Universit\`{a} and Sezione INFN, Padova, Italy
\item \Idef{org31}Dipartimento di Fisica `E.R.~Caianiello' dell'Universit\`{a} and Gruppo Collegato INFN, Salerno, Italy
\item \Idef{org32}Dipartimento di Scienze e Innovazione Tecnologica dell'Universit\`{a} del  Piemonte Orientale and Gruppo Collegato INFN, Alessandria, Italy
\item \Idef{org33}Dipartimento Interateneo di Fisica `M.~Merlin' and Sezione INFN, Bari, Italy
\item \Idef{org34}Division of Experimental High Energy Physics, University of Lund, Lund, Sweden
\item \Idef{org35}Eberhard Karls Universit\"{a}t T\"{u}bingen, T\"{u}bingen, Germany
\item \Idef{org36}European Organization for Nuclear Research (CERN), Geneva, Switzerland
\item \Idef{org37}Excellence Cluster Universe, Technische Universit\"{a}t M\"{u}nchen, Munich, Germany
\item \Idef{org38}Faculty of Engineering, Bergen University College, Bergen, Norway
\item \Idef{org39}Faculty of Mathematics, Physics and Informatics, Comenius University, Bratislava, Slovakia
\item \Idef{org40}Faculty of Nuclear Sciences and Physical Engineering, Czech Technical University in Prague, Prague, Czech Republic
\item \Idef{org41}Faculty of Science, P.J.~\v{S}af\'{a}rik University, Ko\v{s}ice, Slovakia
\item \Idef{org42}Faculty of Technology, Buskerud and Vestfold University College, Vestfold, Norway
\item \Idef{org43}Frankfurt Institute for Advanced Studies, Johann Wolfgang Goethe-Universit\"{a}t Frankfurt, Frankfurt, Germany
\item \Idef{org44}Gangneung-Wonju National University, Gangneung, South Korea
\item \Idef{org45}Gauhati University, Department of Physics, Guwahati, India
\item \Idef{org46}Helsinki Institute of Physics (HIP), Helsinki, Finland
\item \Idef{org47}Hiroshima University, Hiroshima, Japan
\item \Idef{org48}Indian Institute of Technology Bombay (IIT), Mumbai, India
\item \Idef{org49}Indian Institute of Technology Indore, Indore (IITI), India
\item \Idef{org50}Inha University, Incheon, South Korea
\item \Idef{org51}Institut de Physique Nucl\'eaire d'Orsay (IPNO), Universit\'e Paris-Sud, CNRS-IN2P3, Orsay, France
\item \Idef{org52}Institut f\"{u}r Informatik, Johann Wolfgang Goethe-Universit\"{a}t Frankfurt, Frankfurt, Germany
\item \Idef{org53}Institut f\"{u}r Kernphysik, Johann Wolfgang Goethe-Universit\"{a}t Frankfurt, Frankfurt, Germany
\item \Idef{org54}Institut f\"{u}r Kernphysik, Westf\"{a}lische Wilhelms-Universit\"{a}t M\"{u}nster, M\"{u}nster, Germany
\item \Idef{org55}Institut Pluridisciplinaire Hubert Curien (IPHC), Universit\'{e} de Strasbourg, CNRS-IN2P3, Strasbourg, France
\item \Idef{org56}Institute for Nuclear Research, Academy of Sciences, Moscow, Russia
\item \Idef{org57}Institute for Subatomic Physics of Utrecht University, Utrecht, Netherlands
\item \Idef{org58}Institute for Theoretical and Experimental Physics, Moscow, Russia
\item \Idef{org59}Institute of Experimental Physics, Slovak Academy of Sciences, Ko\v{s}ice, Slovakia
\item \Idef{org60}Institute of Physics, Academy of Sciences of the Czech Republic, Prague, Czech Republic
\item \Idef{org61}Institute of Physics, Bhubaneswar, India
\item \Idef{org62}Institute of Space Science (ISS), Bucharest, Romania
\item \Idef{org63}Instituto de Ciencias Nucleares, Universidad Nacional Aut\'{o}noma de M\'{e}xico, Mexico City, Mexico
\item \Idef{org64}Instituto de F\'{\i}sica, Universidad Nacional Aut\'{o}noma de M\'{e}xico, Mexico City, Mexico
\item \Idef{org65}iThemba LABS, National Research Foundation, Somerset West, South Africa
\item \Idef{org66}Joint Institute for Nuclear Research (JINR), Dubna, Russia
\item \Idef{org67}Konkuk University, Seoul, South Korea
\item \Idef{org68}Korea Institute of Science and Technology Information, Daejeon, South Korea
\item \Idef{org69}KTO Karatay University, Konya, Turkey
\item \Idef{org70}Laboratoire de Physique Corpusculaire (LPC), Clermont Universit\'{e}, Universit\'{e} Blaise Pascal, CNRS--IN2P3, Clermont-Ferrand, France
\item \Idef{org71}Laboratoire de Physique Subatomique et de Cosmologie, Universit\'{e} Grenoble-Alpes, CNRS-IN2P3, Grenoble, France
\item \Idef{org72}Laboratori Nazionali di Frascati, INFN, Frascati, Italy
\item \Idef{org73}Laboratori Nazionali di Legnaro, INFN, Legnaro, Italy
\item \Idef{org74}Lawrence Berkeley National Laboratory, Berkeley, California, United States
\item \Idef{org75}Moscow Engineering Physics Institute, Moscow, Russia
\item \Idef{org76}Nagasaki Institute of Applied Science, Nagasaki, Japan
\item \Idef{org77}National Centre for Nuclear Studies, Warsaw, Poland
\item \Idef{org78}National Institute for Physics and Nuclear Engineering, Bucharest, Romania
\item \Idef{org79}National Institute of Science Education and Research, Bhubaneswar, India
\item \Idef{org80}National Research Centre Kurchatov Institute, Moscow, Russia
\item \Idef{org81}Niels Bohr Institute, University of Copenhagen, Copenhagen, Denmark
\item \Idef{org82}Nikhef, Nationaal instituut voor subatomaire fysica, Amsterdam, Netherlands
\item \Idef{org83}Nuclear Physics Group, STFC Daresbury Laboratory, Daresbury, United Kingdom
\item \Idef{org84}Nuclear Physics Institute, Academy of Sciences of the Czech Republic, \v{R}e\v{z} u Prahy, Czech Republic
\item \Idef{org85}Oak Ridge National Laboratory, Oak Ridge, Tennessee, United States
\item \Idef{org86}Petersburg Nuclear Physics Institute, Gatchina, Russia
\item \Idef{org87}Physics Department, Creighton University, Omaha, Nebraska, United States
\item \Idef{org88}Physics Department, Panjab University, Chandigarh, India
\item \Idef{org89}Physics Department, University of Athens, Athens, Greece
\item \Idef{org90}Physics Department, University of Cape Town, Cape Town, South Africa
\item \Idef{org91}Physics Department, University of Jammu, Jammu, India
\item \Idef{org92}Physics Department, University of Rajasthan, Jaipur, India
\item \Idef{org93}Physik Department, Technische Universit\"{a}t M\"{u}nchen, Munich, Germany
\item \Idef{org94}Physikalisches Institut, Ruprecht-Karls-Universit\"{a}t Heidelberg, Heidelberg, Germany
\item \Idef{org95}Purdue University, West Lafayette, Indiana, United States
\item \Idef{org96}Pusan National University, Pusan, South Korea
\item \Idef{org97}Research Division and ExtreMe Matter Institute EMMI, GSI Helmholtzzentrum f\"ur Schwerionenforschung, Darmstadt, Germany
\item \Idef{org98}Rudjer Bo\v{s}kovi\'{c} Institute, Zagreb, Croatia
\item \Idef{org99}Russian Federal Nuclear Center (VNIIEF), Sarov, Russia
\item \Idef{org100}Saha Institute of Nuclear Physics, Kolkata, India
\item \Idef{org101}School of Physics and Astronomy, University of Birmingham, Birmingham, United Kingdom
\item \Idef{org102}Secci\'{o}n F\'{\i}sica, Departamento de Ciencias, Pontificia Universidad Cat\'{o}lica del Per\'{u}, Lima, Peru
\item \Idef{org103}Sezione INFN, Bari, Italy
\item \Idef{org104}Sezione INFN, Bologna, Italy
\item \Idef{org105}Sezione INFN, Cagliari, Italy
\item \Idef{org106}Sezione INFN, Catania, Italy
\item \Idef{org107}Sezione INFN, Padova, Italy
\item \Idef{org108}Sezione INFN, Rome, Italy
\item \Idef{org109}Sezione INFN, Trieste, Italy
\item \Idef{org110}Sezione INFN, Turin, Italy
\item \Idef{org111}SSC IHEP of NRC Kurchatov institute, Protvino, Russia
\item \Idef{org112}Stefan Meyer Institut f\"{u}r Subatomare Physik (SMI), Vienna, Austria
\item \Idef{org113}SUBATECH, Ecole des Mines de Nantes, Universit\'{e} de Nantes, CNRS-IN2P3, Nantes, France
\item \Idef{org114}Suranaree University of Technology, Nakhon Ratchasima, Thailand
\item \Idef{org115}Technical University of Ko\v{s}ice, Ko\v{s}ice, Slovakia
\item \Idef{org116}Technical University of Split FESB, Split, Croatia
\item \Idef{org117}The Henryk Niewodniczanski Institute of Nuclear Physics, Polish Academy of Sciences, Cracow, Poland
\item \Idef{org118}The University of Texas at Austin, Physics Department, Austin, Texas, USA
\item \Idef{org119}Universidad Aut\'{o}noma de Sinaloa, Culiac\'{a}n, Mexico
\item \Idef{org120}Universidade de S\~{a}o Paulo (USP), S\~{a}o Paulo, Brazil
\item \Idef{org121}Universidade Estadual de Campinas (UNICAMP), Campinas, Brazil
\item \Idef{org122}University of Houston, Houston, Texas, United States
\item \Idef{org123}University of Jyv\"{a}skyl\"{a}, Jyv\"{a}skyl\"{a}, Finland
\item \Idef{org124}University of Liverpool, Liverpool, United Kingdom
\item \Idef{org125}University of Tennessee, Knoxville, Tennessee, United States
\item \Idef{org126}University of the Witwatersrand, Johannesburg, South Africa
\item \Idef{org127}University of Tokyo, Tokyo, Japan
\item \Idef{org128}University of Tsukuba, Tsukuba, Japan
\item \Idef{org129}University of Zagreb, Zagreb, Croatia
\item \Idef{org130}Universit\'{e} de Lyon, Universit\'{e} Lyon 1, CNRS/IN2P3, IPN-Lyon, Villeurbanne, France
\item \Idef{org131}V.~Fock Institute for Physics, St. Petersburg State University, St. Petersburg, Russia
\item \Idef{org132}Variable Energy Cyclotron Centre, Kolkata, India
\item \Idef{org133}Warsaw University of Technology, Warsaw, Poland
\item \Idef{org134}Wayne State University, Detroit, Michigan, United States
\item \Idef{org135}Wigner Research Centre for Physics, Hungarian Academy of Sciences, Budapest, Hungary
\item \Idef{org136}Yale University, New Haven, Connecticut, United States
\item \Idef{org137}Yonsei University, Seoul, South Korea
\item \Idef{org138}Zentrum f\"{u}r Technologietransfer und Telekommunikation (ZTT), Fachhochschule Worms, Worms, Germany
\end{Authlist}
\endgroup